\documentclass[a4paper,fleqn,usenatbib]{mnras}
\usepackage{graphicx}
\usepackage{amssymb}
\usepackage{amsmath}
\usepackage[T1]{fontenc}
\usepackage{ae,aecompl}
\usepackage{natbib,times}
\usepackage{lscape}
\newcommand{\vecB}{\bmath B}
\newcommand{\vechatr}{\hat{\bmath r}}
\newcommand{\vechatm}{\hat{\bmath m}}
\newcommand{\vecr}{\bmath r}

\begin{document}
\title[Highly polarized components of pulsars]{Highly polarized components
  of integrated pulse profiles}

\author[P. F. Wang and J. L. Han]
{P. F. Wang\thanks{E-mail: pfwang@nao.cas.cn} and J. L. Han\\
  National Astronomical Observatories, Chinese Academy of
  Sciences.  A20 Datun Road, Chaoyang District, Beijing 100012, China \\
}

\date{Accepted XXX. Received YYY; in original form ZZZ}

\label{firstpage}
\pagerange{\pageref{firstpage}--\pageref{lastpage}}
\maketitle

\begin{abstract}
Highly polarized components of pulse profiles are investigated by
analyzing observational data and simulating the emission
processes. The highly polarized components appear at the leading or
trailing part of a pulse profile, which preferably have a flat
spectrum and a flat polarization angle curve compared with the low
polarized components. By considering the emission processes and
propagation effects, we simulate the distributions of wave modes and
fractional linear polarization within the entire pulsar emission
beam. We show that the highly polarized components can appear at the
leading, central, and/or trailing parts of pulse profiles in the
models, depending on pulsar geometry. The depolarization is caused by
orthogonal modes or scattering. When a sight line cuts across pulsar
emission beam with a small impact angle, the detected highly polarized
component will be of the O mode, and have a flat polarization angle
curve and/or a flat spectrum as observed. Otherwise, the highly
polarized component will be of the X mode and have a steep
polarization angle curve.

\end{abstract}


\section{INTRODUCTION}

Integrated pulse profiles are obtained by integrating tens of
thousands of individual pulses. Features of pulse profiles have been
investigated to understand the geometry and physical processes within
pulsar magnetosphere \citep[e.g.][]{ran83, lm88, kwj+94,
  nsk+15}. Integrated pulse profiles generally comprise several
components, and are characterized by diverse polarization features,
including prominent linear polarization, `S'-shaped polarization angle
curves, single sign or sign reversals of circular polarization. Some
integrated pulse profiles are highly polarized for the whole pulse,
even 100\% polarized such as PSR B1259-63 and B1823-13. These pulsars
are generally young and have very high spin-down luminosity $\dot{E}$
and flat spectrum \citep{qml+95, vkk98, cmk01, wj08}. Some pulsars
have highly linearly polarized leading or trailing components, for
example, the leading components of PSRs B0355+54 and B0450+55
\citep{lm88, vx97, gl98}, and the trailing components of PSRs B1650-38
and B1931+24 \citep{kjm05, hdv+09}. \citet{vkk98} noticed that the
highly polarized leading component of PSR B0355+54 has a flat spectrum
and becomes increasingly prominent at higher frequencies. The highly
polarized trailing component of PSR B2224+65 has a flat polarization
angle curve \citep{mr11}.

\begin{figure*}
  \centering
  \begin{tabular}{c}
    \includegraphics[angle=0, width=0.8\textwidth] {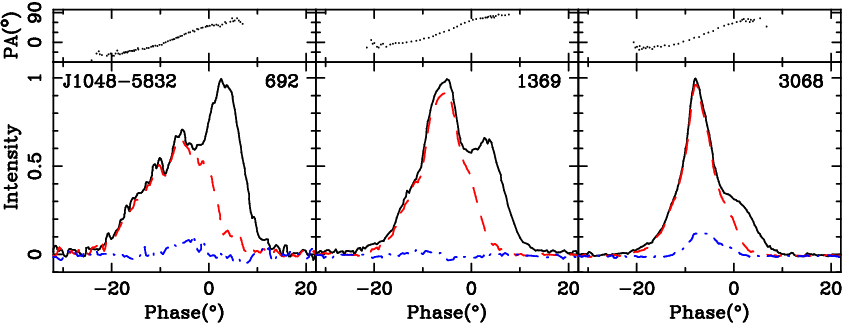} \\
    \includegraphics[angle=0, width=0.8\textwidth] {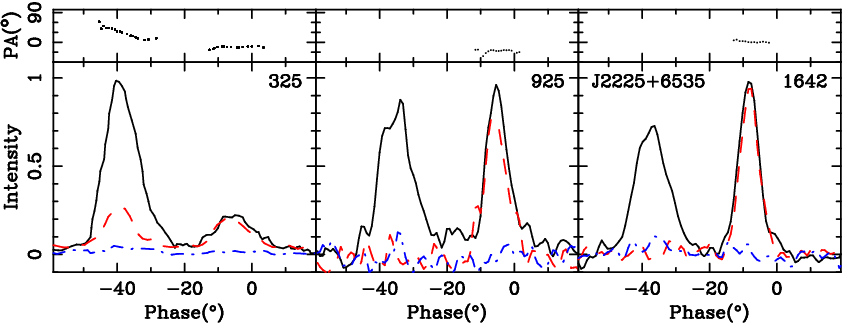} \\
  \end{tabular}
  \caption{Highly polarized leading component of PSR J1048-5832 and
    trailing component of J2225+6535 at three frequencies to show
    their frequency evolution. The solid lines stand for the total
    intensities, the dashed and dashed-dotted lines represent the
    linear and circular polarizations, respectively. The position
    angle curves are shown by dotted lines at the upper part of each
    panel. The polarization data are collected from literature as
    listed in Table \ref{high_L}. }
  \label{fig:profiles}
\end{figure*}

Theoretical efforts have been made to understand various polarization
features. In general, pulsar polarizations are closely related to the
emission processes of the relativistic particles streaming along the
curved magnetic field lines \citep[e.g.][]{bcw91,wwh12}, the
propagation effects within pulsar magnetosphere
\citep[e.g.][]{ba86,wlh10,bp12}, and the scattering within the
interstellar medium \citep{lh03}. However, these investigations have
been conducted separately on each aspect, rarely done
jointly. Curvature radiation which serves as one of the most probable
mechanisms for pulsar emission can produce highly polarized emission
\citep{gan10,wwh12}. Propagation effects are succeeded in
demonstrating the interaction of the ordinary (O) and extra-ordinary
(X) modes within pulsar magnetosphere and can lead to diverse
depolarization features \citep{cr79,ba86,wlh10,bp12}, though initial
ratios for both modes are uncertain. Propagation effects within the
interstellar medium need to be investigated further. Recently, we
investigated the emission processes jointly with propagation effects
\citep{wwh14, wwh15}, which provides us a new opportunity to
understand the highly polarized components, because distributions of
the X-mode and O-mode within pulsar magnetosphere are related to the
depolarization across pulsar emission beam by considering the
refraction and corotation effects. Emission can be highly depolarized
in some beam regions where both modes have comparable intensities, but
dominated by one mode in other regions and hence the resulting profile
can be highly polarized.

In this paper, we summarize observations for highly polarized
components of integrated pulse profiles in literature and then
theoretically explain them by modeling emission and propagation
processes. In Section 2, we analyze various features for highly
polarized components of observed pulsar profiles. In Section 3, we
simulate polarized pulsar beams and pulse profiles by considering the
emission processes and propagation effects. Discussions and
conclusions are given in Section 4.

\begin{table*}
    \caption{Highly polarized components of 78 pulsars in literature.  }
    \label{high_L}
    \tabcolsep 1mm 
    \begin{tabular}{llrrll}
      \hline
      \hline
      PSR Jname  &  Bname  & Period  & DM     & Polarization Features &  References           \\
                 &         &  (s)    & ($\rm cm^{-3} pc$)&                       &            \\
      \hline
      J0014+4746 & B0011+47 & 1.24069 & 30.8  & Leading                         & 31, 70, 86                 \\
      J0358+5413 & B0355+54 & 0.15638 & 57.1  & Leading, Flat Spec., Orth. Modes & 8, 9, 16, 20, 24, 26, 30, 31, 34, 77, 81     \\
      J0454+5543 & B0450+55 & 0.34072 & 14.5  & Leading, Flat Spec., Flat PA  & 16, 20, 30, 31, 81         \\
      J0814+7429 & B0809+74 & 1.29224 & 5.7   & Leading, Orth. Modes             & 1, 9, 30, 31, 44, 58, 78, 88\\
      J0942-5657 & B0941-56 & 0.80812 & 159.7  & Leading                        & 23, 33, 69, 90             \\
      J0954-5430 &          & 0.47283 & 200.3  & Leading                        & 69, 90                     \\
      J1048-5832 & B1046-58 & 0.12367 & 129.1  & Leading, Flat Spec.            & 23, 56, 59, 69, 76, 89, 90 \\
      J1057-5226IP& B1055-52& 0.19710 & 30.1   & Leading, Flat Spec.            & 4, 6, 12, 16, 29, 69, 72, 77, 89, 90       \\
      J1112-6103 &          & 0.06496 & 599.1  & Leading, Scattering            & 69, 89, 90                \\
      J1341-6220 & B1338-62 & 0.19333 & 717.3  & Leading, Scattering            & 23, 39, 59,  60, 69, 90   \\
      J1410-6132 &          & 0.05005 & 960.0  & Leading, Scattering            & 68, 69, 89, 90            \\
      J1453-6413 & B1449-64 & 0.17948 & 71.0   & Leading, Flat PA, Orth. Modes & 5, 6, 7, 29, 54, 59, 69, 71, 81, 90   \\
      J1730-3350 & B1727-33 & 0.13946 & 259.0  & Leading, Scattering            & 31, 39, 59, 69, 89, 90     \\
      J1805+0306 & B1802+03 & 0.21871 & 80.8   & Leading, Flat PA             & 31, 38                      \\
      J1823-3106 & B1820-31 & 0.28405 & 50.2   & Leading                        & 21, 31                     \\
      J1825-0935MP& B1822-09& 0.76900 & 19.3   & Leading, Flat Spec., Orth. Modes& 7, 9, 24, 29, 30, 31, 49, 61, 63, 65, 73, 77  \\
      J1844+1454 & B1842+14 & 0.37546 & 41.4   & Leading, Flat Spec., Flat PA & 17, 31, 38, 54, 65, 74, 77, 81, 90     \\
      J1849+2423 &          & 0.27564 & 62.2   & Leading                        & 70                          \\
      J1937+2544 & B1935+25 & 0.20098 & 53.2   & Leading, Flat PA             & 31, 38, 63, 70, 81, 90      \\
      J2008+2513 &          & 0.58919 & 60.5   & Leading, Orth. Modes            & 70                          \\

      \hline

      J0601-0527 & B0559-05& 0.39596 & 80.5    & Trailing, Flat Spec., Flat PA & 20, 23, 31, 34, 54, 69, 90  \\
      J0922+0638 &B0919+06 & 0.43062 & 27.2    & Trailing, Flat Spec., Orth. Modes & 13, 19, 24, 29--31, 38, 49, 54, 59     \\
                 &         &         &         &       & 62, 65, 74, 81, 90  \\
      J1401-6357 & B1358-63& 0.84278 & 98.0    & Trailing                         & 21, 23, 29                  \\
      J1539-5626 & B1535-56& 0.24339 & 175.8   & Trailing, Flat Spec., Flat PA  & 23, 56, 59, 69, 90          \\
      J1548-5607 &         & 0.17093 & 315.5   & Trailing                         & 69, 90                      \\
      J1653-3838 &B1650-38 & 0.30503 & 207.2   & Trailing                         & 56, 69, 90                  \\
      J1739-1313 &         & 1.21569 & 58.2    & Trailing                         & 69, 90                      \\
      J1808-3249 &         & 0.36491 & 147.3   & Trailing                         & 56, 69, 90                  \\
      J1933+2421 &B1931+24 & 0.81369 & 106.0   & Trailing                         & 31, 70                      \\
      J2013+3845 & B2011+38& 0.23019 & 238.2   & Trailing, Flat PA              & 31, 81                      \\
      J2225+6535 &B2224+65 & 0.68254 & 36.0    & Trailing, Flat Spec., Flat PA  & 9, 16, 31, 77, 86, 88       \\

      \hline

      J0737-3039A&         & 0.02269 & 48.9    & MSP, Leading\&Trailing  &  47, 55, 57, 82      \\
                 &         &         &         & Flat PA, Orth. Mod      & \\
      J1012+5307 &         & 0.00525 & 9.0     & MSP, Trailing, Flat PA         & 35, 37, 70, 88              \\
      J1022+1001 &         & 0.01645 & 10.2    & MSP, Trailing                    & 35--37, 46, 48, 52, 80, 84, 85, 87, 88  \\
      J1300+1240 &B1257+12 & 0.00621 & 10.1    & MSP, Leading                     & 35, 70                    \\

      \hline

      J0108-1431 &          & 0.80756 & 2.3    & Whole                          & 33, 69, 90                  \\
      J0134-2937 &          & 0.13696 &	21.8   & Whole                          & 33, 65, 69, 71, 90          \\
      J0139+5814 & B0136+57 & 0.27245 &	73.7   & Whole                          & 16, 20, 30, 31, 81, 86, 88  \\
      J0538+2817 &          & 0.14315 &	39.5   & Whole                          & 30, 81                      \\
      J0543+2329 & B0540+23 & 0.24597 & 77.7   & Whole, Pol. dec. with freq.    & 9, 10, 17, 19, 24, 30, 31, 38, 49       \\
                 &         &         &         &       & 53, 63, 65, 69, 74, 77, 81, 90         \\
      J0614+2229 & B0611+22 & 0.33495 &	96.9   & Whole, Strong CP               & 10, 16, 19, 31, 34, 38, 53, 63, 65, 69, 74   \\
      J0630-2834 & B0628-28 & 1.24441 & 34.4   & Whole                          & 5, 6, 7, 9, 16, 29, 31, 34, 54, 59, 65, 69, 81, 90  \\
      J0631+1036 &          & 0.28780 & 125.3  & Whole                          & 27, 69, 75, 89, 90          \\
      J0659+1414 & B0656+14 & 0.38489 & 13.9   & Whole                          & 17, 31, 38, 40, 53, 59, 63, 64, 69, 74   \\
                 &         &         &         &       & 75, 81, 89, 90  \\
      J0742-2822 & B0740-28 & 0.16676 & 73.7   & Whole                          & 6, 7, 9, 16, 24, 29--31, 33, 49, 54, 59   \\
                 &         &         &         &       & 61, 69, 71, 75, 77, 81, 83, 89, 90                                 \\
      J0835-4510 & B0833-45 & 0.08932 & 67.9   & Whole, Strong CP               & 2, 3, 5--7, 11, 29, 43, 54, 59, 61        \\
                 &         &         &         &       & 69, 71, 76, 81, 89, 90 \\ 
      J0901-4624 &          & 0.44199 & 198.8  & Whole, Strong CP               & 69, 90                       \\
      J0905-5127 &          & 0.34628 & 196.4  & Whole                          & 69, 90                       \\
      J0908-4913 & B0906-49 & 0.10675 & 180.3  & Whole, Inter Pulse             & 21, 23, 32, 59, 66, 69, 76, 89, 90         \\
      J1015-5719 &          & 0.13988 & 278.7  & Whole                          & 60, 69, 90                   \\
      J1028-5819 &          & 0.09140 & 96.5   & Whole                          & 67, 69, 89                   \\
      J1057-5226MP& B1055-52& 0.19710 & 30.1   & Whole                          & 4, 6, 12, 16, 29, 69, 72,77, 89, 90   \\
      J1105-6107 &          & 0.06319 & 271.0  & Whole                          & 39, 60, 69, 89, 90           \\
      \hline
    \end{tabular}
\end{table*}
\begin{table*}
  \addtocounter{table}{-1}
   \caption{ -- continued.  }
    \tabcolsep 1mm
    \begin{tabular}{lllrll}
      \hline
      J1119-6127 &          & 0.40796 & 707.4  & Whole                          & 45, 60, 69, 79, 89, 90       \\
      J1302-6350 & B1259-63 & 0.04776 & 146.7  & Whole, Strong CP               & 22, 25, 42, 56, 59, 69, 90   \\
      J1321+8323 & B1322+83 & 0.67003 & 13.3   & Whole                          & 31, 70, 86                   \\
      J1359-6038 & B1356-60 & 0.12750 & 293.7  & Whole, Strong CP               & 21, 29, 33, 59, 61, 69, 90   \\
      J1420-6048 &          & 0.06817 & 358.8  & Whole, Strong CP               & 41, 60, 69, 75, 89, 90       \\
      J1614-5048 & B1610-50 & 0.23169 & 582.8  & Whole, Scattering, Strong CP   & 23, 56, 69, 90               \\
      J1637-4553 & B1634-45 & 0.11877 & 193.2  & Whole                          & 56, 69, 90                   \\
      J1702-4128 &          & 0.18213 & 367.1  & Whole                          & 69, 89, 90                   \\
      J1705-1906IP& B1702-19& 0.29898 & 22.9   & Whole, Strong CP, Inter-pulse         & 15, 16, 29, 31, 34, 49, 65, 69, 90  \\
      J1705-3950 &          & 0.31894 & 207.1  & Whole, Strong CP               & 69, 90                       \\
      J1709-4429 & B1706-44 & 0.10245 & 75.6   & Whole, Strong CP               & 23, 54, 56, 59, 69, 89, 90   \\
      J1718-3825 &          & 0.07466 & 247.4  & Whole                          & 69, 75, 89, 90               \\
      J1733-3716 & B1730-37 & 0.33758 & 153.5  & Whole                          & 56, 69, 90                   \\
      J1740-3015 & B1737-30 & 0.60688 & 152.1  & Whole, Strong CP               & 21, 23, 31, 34, 59, 69, 76, 90 \\
      J1801-2451 & B1757-24 & 0.12491 & 289.0  & Whole, Strong CP               & 31, 63, 69, 81, 86, 89, 90   \\
      J1803-2137 & B1800-21 & 0.13366 & 233.9  & Whole, Strong CP               & 21, 31, 34, 69, 86, 90       \\
      J1809-1917 &          & 0.08274 & 197.1  & Whole, Strong CP               & 69, 90                       \\
      J1826-1334 & B1823-13 & 0.10148 & 231.0  & Whole, Strong CP               & 31, 34, 69, 90               \\
      J1830-1059 & B1828-11 & 0.40504 & 161.5  & Whole, Strong CP               & 31, 69, 90                   \\
      J1841-0345 &          & 0.20406 & 194.3  & Whole                          & 69, 90                       \\
      J1841-0425 & B1838-04 & 0.18614 & 325.4  & Whole                          & 31, 63, 69, 90               \\
      J1850+1335 & B1848+13 & 0.34558 & 60.1   & Whole                          & 31, 38, 63, 65, 81, 90       \\
      J1915+1009 & B1913+10 & 0.40454 & 241.6  & Whole, Strong CP               & 17, 31, 34, 38, 63, 81, 90   \\
      J1926+1648 & B1924+16 & 0.57982 & 176.8  & Whole                          & 10, 17, 19, 31, 38, 77       \\
      J1932+1059 & B1929+10 & 0.22651 & 3.1    & Whole                          & 2, 9, 10, 13, 14, 16--20, 24, 26, 28--31, 37, 38  \\
                 &          &         &        &                                  & 40, 49--51, 53, 54, 70, 74, 81, 86, 88, 90      \\

      \hline
    \end{tabular}
    \parbox{180mm} { Notes. Leading: Highly polarized (larger than 70\%) leading components;  Trailing:
      highly polarized trailing components; MSP: millisecond pulsars with highly polarized
      leading and/or trailing components; Whole: highly polarized for the whole pulse profiles;
      Flat Spec.: flat spectrum; Flat PA: flat polarization angle curves; Orth. Modes: orthogonal
modes; Strong CP: strong circular polarization. References: (1)
      \citet{lsg71} at 0.151, 0.24, 0.408 GHz; (2) \citet{man71} at
      0.392, 1.665 GHz; (3) \citet{kmr74} at 4.83 GHz; (4)
      \citet{mha+76} at 1.4 GHz; (5) \citet{hma+77} at 0.338, 0.4 GHz;
      (6) \citet{mhm+78} at 0.631, 0.649 GHz; (7) \citet{mhm80} at
      1.612 GHz; (8) \citet{msf+80} at 2.65 GHz; (9) \citet{mgs+81} at
      1.72, 2.65, 4.85, 8.7 GHz; (10) \citet{rb81} at 0.43 GHz; (11)
      \citet{kd83} at 2.295 GHz; (12) \citet{ran83} at 0.17, 0.631
      GHz; (13) \citet{scr+84} at 1.404 GHz; (14) \citet{scw+84} at
      0.8 GHz; (15) \citet{blh+88} at 0.408 GHz; (16) \citet{lm88} at
      0.408, 0.415, 0.43, 0.611, 0.64, 1.42 GHz; (17) \citet{rsw89} at
      1.4 GHz; (18) \citet{ph90} at 0.43, 1.665 GHz; (19)
      \citet{bcw91} at 0.43, 1.418 GHz; (20) \citet{xrs+91} at 1.72
      GHz; (21) \citet{wml+93} at 1.56 GHz; (22) \citet{mj95} at 1.52,
      4.68 GHz; (23) \citet{qml+95} at 0.66, 1.411, 1.44 GHz; (24)
      \citet{xsg+95} at 10.55 GHz; (25) \citet{jml+96} at 4.8 GHz;
      (26) \citet{xkj+96} at 32.0 GHz; (27) \citet{zcw+96} at 1.418,
      1.665, 2.38 GHz; (28) \citet{rr97} at 0.43, 1.414 GHz; (29)
      \citet{vdh+97} at 0.8, 0.95 GHz; (30) \citet{vx97} at 1.41,
      1.71, 4.85, 10.55 GHz; (31) \citet{gl98} at 0.23, 0.4, 0.6,
      0.92, 1.4, 1.6 GHz; (32) \citet{gsf+98} at 1.3 GHz; (33)
      \citet{mhq98} at 0.435, 0.66 GHz; (34) \citet{vkk98} at 4.85,
      10.55 GHz; (35) \citet{xkj+98} at 1.41 GHz; (36) \citet{kxc+99}
      at 1.41 GHz; (37) \citet{stc99} at 0.41, 0.61, 1.414 GHz; (38)
      \citet{wcl+99} at 1.418 GHz; (39) \citet{cmk01} at 0.661, 1.351
      GHz; (40) \citet{ew01} at 1.418 GHz; (41) \citet{rrj01} at 1.517
      GHz; (42) \citet{cjm+02} at 1.4 GHz; (43) \citet{kjv02} at 2.3
      GHz; (44) \citet{rrs+02} at 0.328, 1.365 GHz; (45) \citet{ck03}
      at 1.366, 2.496 GHz; (46) \citet{rk03} at 1.42 GHz; (47)
      \citet{drb+04} at 0.82 GHz; (48) \citet{hbo04} at 1.341 GHz;
      (49) \citet{kj04} at 1.4, 4.85 GHz; (50) \citet{mc04} at 1.404
      GHz; (51) \citet{mr04} at 0.43, 1.17 GHz; (52) \citet{ovh+04} at
      1.373 GHz; (53) \citet{wck+04} at 0.43 GHz; (54) \citet{jhv+05}
      at 1.4 GHz; (55) \citet{hbo05} at 0.685, 1.373 GHz; (56)
      \citet{kjm05} at 3.1 GHz; (57) \citet{rdk+05} at 0.82 GHz; (58)
      \citet{rrs05} at 0.112, 0.328 GHz; (59) \citet{jkw06} at 8.4
      GHz; (60) \citet{jw06} at 1.369, 3.1 GHz; (61) \citet{kj06} at
      1.375, 3.1 GHz; (62) \citet{rrw06} at 0.327, 1.425 GHz; (63)
      \citet{jkk+07} at 0.691, 1.374, 3.1 GHz; (64) \citet{ran07} at
      1.525 GHz; (65) \citet{jkm+08} at 0.243, 0.322, 0.69, 1.4, 3.1
      GHz; (66) \citet{kj08} at 1.4, 3.1, 8.6 GHz; (67) \citet{kjk+08}
      at 1.37, 3.087 GHz; (68) \citet{ojk+08} at 3.1, 6.2 GHz; (69)
      \citet{wj08} at 1.5, 3.0 GHz; (70) \citet{hdv+09} at 0.774 GHz;
      (71) \citet{nkk+09} at 1.369, 1.375 GHz; (72) \citet{ww09} at
      1.369 GHz; (73) \citet{bmr10} at 0.325 GHz; (74) \citet{hr10} at
      0.0492, 0.132, 0.43, 1.404 GHz; (75) \citet{waa+10} at 1.369
      GHz; (76) \citet{kjl+11} at 17,24 GHz; (77) \citet{mr11} at
      0.325 GHz; (78) \citet{rd11} at 0.82 GHz; (79) \citet{wje11} at
      1.5 GHz; (80) \citet{ymv+11} at 1.369 GHz; (81) \citet{nkc+12}
      at 1.4, 2.7, 3.1, 4.85 GHz; (82) \citet{gkj+13} at 1.4 GHz; (83)
      \citet{ksj13} at 1.369, 3.1 GHz; (84) \citet{van13} at 1.341
      GHz; (85) \citet{dhm+15} at 0.6, 1.5, 3.0 GHz; (86)
      \citet{fdr15} at 1.5 GHz; (87) \citet{lkl+15} at 1.3 GHz; (88)
      \citet{nsk+15} at 0.15 GHz; (89) \citet{rwj15} at 1.5, 3.1, 6.0
      GHz; (90) Johnston
      et.al. http://www.atnf.csiro.au/people/joh414/ppdata/.  }
\end{table*}

\section{Observational features for highly polarized pulse components}

The highly polarized components of integrated pulse profiles exhibit
diverse polarization features. To demonstrate the properties, a sample
of 78 pulsars is collected from literatures, as listed in
Table~\ref{high_L}. Among them, 20 pulsars have highly polarized
leading components, 11 pulsars have highly polarized trailing
components, four millisecond pulsars have both highly polarized
leading and/or trailing components, and 43 pulsars are highly
polarized for the whole pulse profile. The fractional linear
polarization is larger than 70\% for highly polarized components or
the whole profile for these pulsars at more than one frequency.

\subsection{Flat spectra of highly polarized components}

Multi-frequency observations demonstrate that pulsar flux density
generally decreases with frequency, following a power-law spectrum
\citep[e.g.][]{sie73}. Different components for a given pulsar could
evolve differently with frequency. For example, the relative spectra
for the leading and trailing components are diverse for the conal
double pulsars \citep{whq01}. The highly polarized components also
show frequency evolution. Fig.~\ref{fig:profiles} shows the polarized
pulse profiles at three frequencies for two pulsars, J1048-5832 and
J2225+6535. PSR J1048-5832 exhibits highly polarized leading component
with polarization degree approaching 100\%. At 692 MHz, the peak
intensity of the highly polarized leading component is weaker than the
low polarized trailing component. As observation frequency increases,
the highly polarized leading component gradually dominates, as shown
by the profiles of 1369 and 3068 MHz. Similar features have been seen
from PSRs J0358+5413, J0454+5543, J1057-5226IP, J1825-0935MP and
J1844+1454. In contrast, PSR J2225+6535 is an example for highly
polarized trailing component, which becomes dominant as observation
frequency increases. Similar cases can be found from PSRs J0601-0527,
J0922+0638 and J1539-5626.

\begin{figure}
  \centering
  \begin{tabular}{cc}
    \includegraphics[angle=0, width=0.222\textwidth] {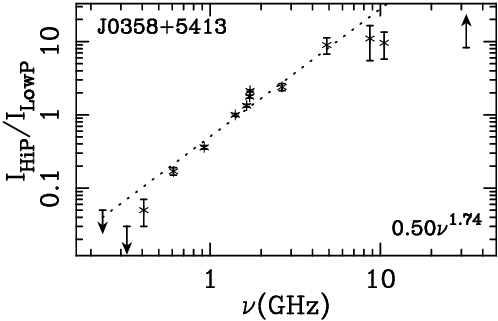} &
    \includegraphics[angle=0, width=0.222\textwidth] {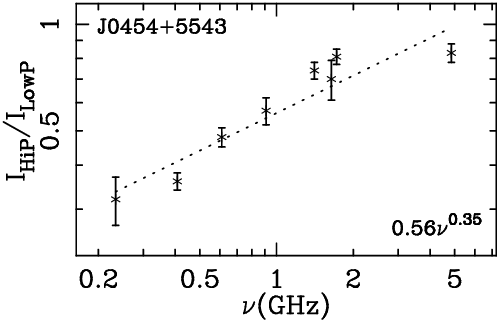} \\
    \includegraphics[angle=0, width=0.222\textwidth] {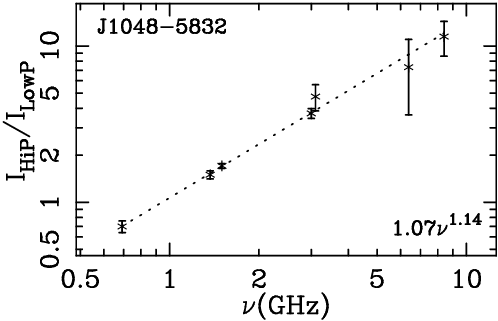} &
    \includegraphics[angle=0, width=0.222\textwidth] {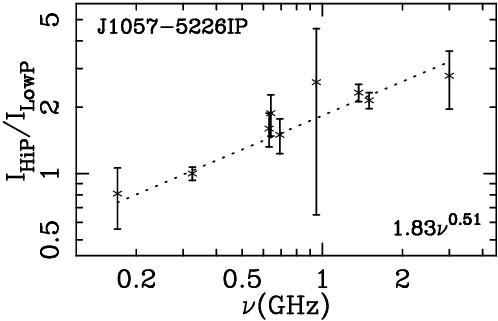} \\
    \includegraphics[angle=0, width=0.222\textwidth] {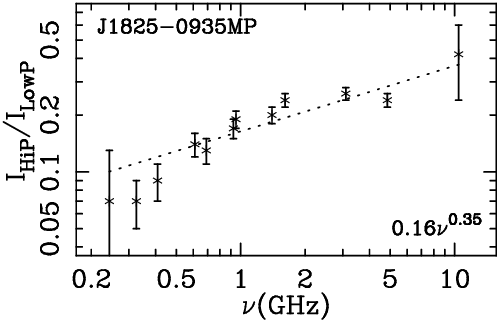} &
    \includegraphics[angle=0, width=0.222\textwidth] {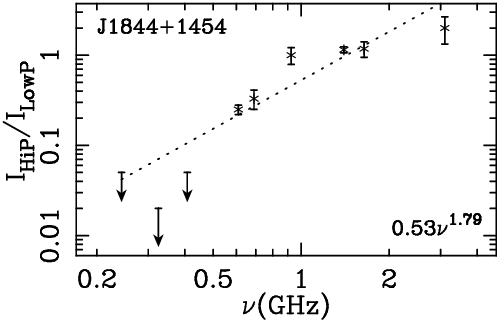} \\
  \end{tabular}
  \caption{The frequency evolution for the peak intensity ratio of the
    highly polarized leading components for six pulsars with respect
    to the low polarized trailing ones. The intensity ratios are
    listed in Table~\ref{table:leading_ratio}, which can be described
    by a power-law as $\rm I_{HiP}/I_{LowP} \sim a\nu^k$. }
  \label{fig:leading_ratio}
\end{figure}

\begin{figure}
  \centering
  \begin{tabular}{cc}
    \includegraphics[angle=0, width=0.222\textwidth] {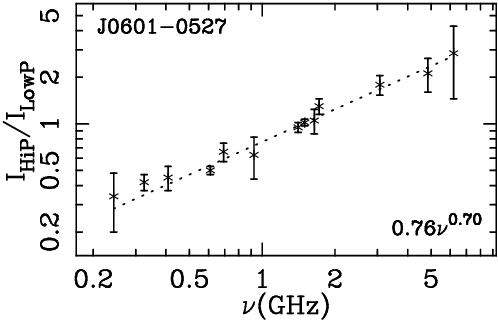} &
    \includegraphics[angle=0, width=0.222\textwidth] {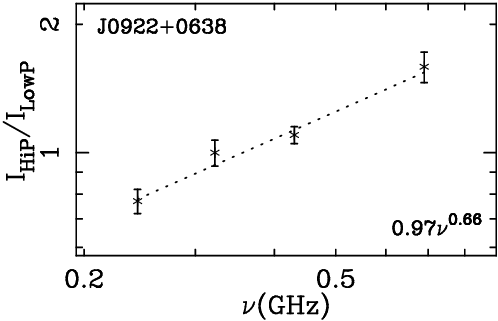} \\
    \includegraphics[angle=0, width=0.222\textwidth] {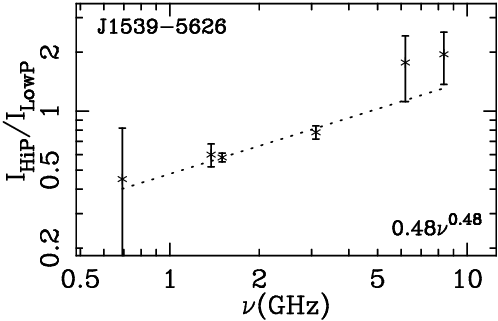} &
    \includegraphics[angle=0, width=0.222\textwidth] {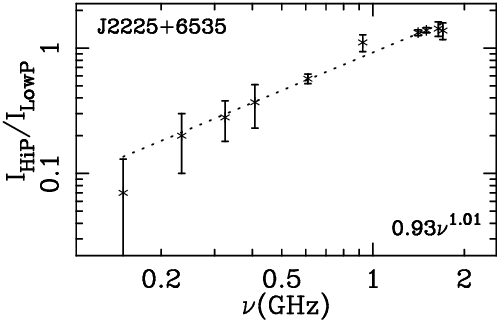} \\
  \end{tabular}
  \caption{Same as Fig.~\ref{fig:leading_ratio} but for the highly
    polarized trailing components for four pulsars. Data are listed in
    Table~\ref{table:trailing_ratio}. }
  \label{fig:trailing_ratio}
\end{figure}

Figs.~\ref{fig:leading_ratio} and \ref{fig:trailing_ratio}
quantitatively demonstrate the frequency evolution of the peak
intensity ratios, $\rm I_{HiP}/I_{LowP}$, of the highly polarized
components with respect to the low polarized ones at a series of
frequencies, see data in Tables \ref{table:leading_ratio} and
\ref{table:trailing_ratio} in the Appendix. Clearly, $\rm
I_{HiP}/I_{LowP}$ generally increases with frequency for highly
polarized leading or trailing components, and can be described by a
power-law, though the power-law indices vary from 0.35 to 1.79 for
different pulsars. We conclude that the highly polarized components
exhibit a flatter spectrum than the low polarized components,
regardless of its location at the leading or trailing phase.

\subsection{Polarization angle curves of highly polarized components}

\begin{table}
\begin{footnotesize}
\caption{Gradients of polarization curves for the highly polarized and
  low polarized components. References are numbered in Table A1.}
\label{table:PA_gradient}
\tabcolsep 1.5mm
    \scriptsize
    \begin{tabular}{lrlrrl}
      \hline
      \hline
      PSR  & Hi. Pol. Comp. &Freq. &$\Delta PA/\Delta\phi$ & $\Delta PA/\Delta \phi$ & Ref.  \\
           &       &(GHz) &  High Pol. Comp.      & Low Pol. Comp.          &       \\
      \hline
      J0454+5543 & Leading  & 1.41  & $-0.6\pm0.2$ & $-7.8\pm0.3$ & 30  \\
      J1453-6413 & Leading  & 1.4   & $0.0\pm0.2$  & $8.0\pm0.3$  & 81 \\
      J1805+0306 & Leading  & 1.4   & $-1.2\pm0.5$ & $13.1\pm0.6$ & 38 \\
      J1844+1454 & Leading  & 1.4   & $2.4\pm1.1$  & $17.0\pm2.1$ & 54  \\
      J1937+2544 & Leading  & 0.774 & $-1.5\pm0.3$ & $-7.6\pm0.6$ & 70 \\
      J0601-0527 & Trailing & 0.692 & $2.6\pm0.6$  & $5.6\pm1.0$  & 90   \\
      J2225+6535 & Trailing & 0.325 & $-0.1\pm0.8$ & $-3.7\pm0.7$ & 77  \\
      \hline
      J0358+5413 & Leading  & 1.408 & $-1.4\pm0.2$ & Orth. Modes  & 31   \\
      J0814+7429 & Leading  & 1.41  & $-0.6\pm0.2$ & Orth. Modes  & 30   \\
      J1825-0935 & Leading  & 0.691 & $3.7\pm0.2$  & Orth. Modes  & 63 \\
      J2008+2513 & Leading  & 0.774 & $1.4\pm0.7$  & Orth. Modes  & 70  \\
      J0922+0638 & Trailing & 0.692 & $5.2\pm0.4$  & Orth. Modes  & 90   \\
      \hline
      J1112-6103 & Leading  & 1.5   & $-0.6\pm0.1$ & Scattering   & 69 \\
                 &          & 3.0   & $-5.0\pm0.1$ &  -           & 69  \\
      J1341-6220 & Leading  & 1.5   & $0.6\pm0.1$  & Scattering   & 60 \\
                 &          & 3.0   & $7.1\pm0.2$  &  -           & 60 \\
      J1410-6132 & Leading  & 1.5   & $0.0\pm0.1$  & Scattering   & 69 \\
                 &          & 3.1   & $4.0\pm0.2$  &  -           & 68 \\
      J1730-3350 & Leading  & 1.5   & $-1.5\pm0.1$ & Scattering   & 69 \\
                 &          & 3.0   & $-5.4\pm0.3$ &  -           & 69 \\
      \hline
      J0014+4746 & Leading  & 0.774 & $-1.0\pm0.1$ & $-1.4\pm0.1$ & 70 \\
      J0954-5430 & Leading  & 1.4   & $12.1\pm2.1$ & $10.8\pm1.7$ & 90  \\
      J1057-5226IP& Leading & 1.377 & $0.6\pm0.5$  & $1.2\pm1.3$  & 90  \\
      J0942-5657 & Leading  & 1.5   & $14.2\pm0.4$ & Mixed        & 69  \\
      J1048-5832 & Leading  & 1.369 & $4.3\pm0.2$  & Mixed        & 90  \\
      J1823-3106 & Leading  & 1.4   & $-4.5\pm0.3$ & Mixed        & 31 \\
      J1401-6357 & Trailing & 0.955 & $8.1\pm1.4$  & Mixed        & 29 \\
      J1739-1313 & Trailing & 1.377 & $8.9\pm1.5$  & Mixed        & 90   \\
      J2013+3845 & Trailing & 1.408 & $-1.0\pm0.2$ & Mixed        & 31   \\
      J1849+2423 & Leading  & 0.774 & $-1.2\pm0.2$ & Weak Pol.    & 70 \\
      J1539-5626 & Trailing & 1.5   & $0.0\pm0.1$  & Weak Pol.    & 69   \\
      J1548-5607 & Trailing & 1.4   & $2.4\pm0.3$  & Weak Pol.    & 90   \\
      J1653-3838 & Trailing & 1.377 & $3.4\pm1.2$  & Weak Pol.    & 90   \\
      J1808-3249 & Trailing & 1.377 & $-8.2\pm1.6$ & Weak Pol.    & 90   \\
      J1933+2421 & Trailing & 0.774 & $9.4\pm0.3$  & Weak Pol.    & 70   \\
      \hline
      J0737-3039A& MSP-Leading  & 1.4   & $0.0\pm0.2$  & Orth. Modes &  82 \\
                 & MSP-Trailing & 1.4   & $0.0\pm0.3$  & -           &  82  \\
      J1012+5307 & MSP-Trailing & 0.774 & $-0.4\pm0.1$ & Mixed       &  70 \\
      J1022+1001 & MSP-Trailing & 1.3   & $3.3\pm0.2$  & $4.4\pm0.1$ &  87  \\
      J1300+1240 & MSP-Leading  & 0.774 & $0.4\pm0.3$  & Weak Pol.   &  70   \\
      \hline
     \end{tabular}
\parbox{85mm}{ Note: Gradients of polarization angle curves for the
  low polarized components are hard to determine due to various
  reasons as listed in the fifth column. }
\end{footnotesize}
\end{table}

Highly polarized components differ from low polarized components also
in polarization angle curves. Table~\ref{table:PA_gradient} summarizes
the gradients of polarization angle curves for 35 pulsars extracted
from Table~\ref{high_L}. The highly polarized components generally
have flat polarization angle curves. For example, the gradient of
polarization angle curve for the highly polarized trailing component
of PSR J2225+6535 in Fig.~\ref{fig:profiles} approximates to be -0.1
at 325MHz, but it is -3.7 for the low polarized leading component
\citep{mr11}. The gradient is 2.4 for the highly polarized leading
component of PSR J1844+1454 at 1.4GHz, but 17.0 for low polarized
trailing component \citep{jhv+05}. The large difference for the
gradients can also be found for PSRs J0454+5543, J1453-6413,
J1805+0306, J1937+2544 and J0601-0527, as listed in
Table~\ref{table:PA_gradient}. It implies that the highly polarized
emission of these pulsars might be generated from the beam regions
well away from the magnetic meridional plane.

However, the highly polarized emission of some pulsars might also be
produced near the meridional plane, e.g. J0942-5657 and
J1933+2421. Both of them have very steep polarization angle curves
with gradients of 14.2 and 9.4 for the highly polarized
components. Gradients for low polarized components of many pulsars are
hard to determine due to various reasons as noted in the fifth column
of Table~\ref{table:PA_gradient}.

\begin{figure}
  \centering
  \includegraphics[angle=0, width=0.4\textwidth] {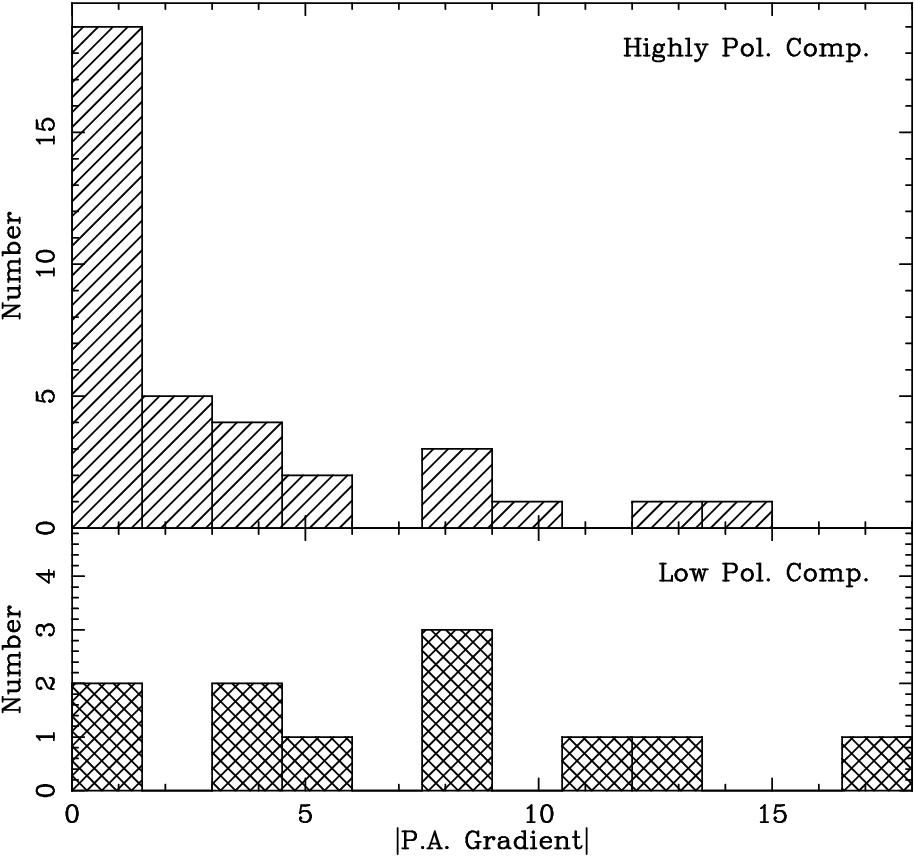}
  \caption{Histograms for absolute values of gradients of the
    polarization angle curves for highly polarized and low polarized
    components. The gradient values are listed in
    Table~\ref{table:PA_gradient}. }
  \label{fig:PA_gradient}
\end{figure}

As shown in Fig.~\ref{fig:PA_gradient}, the gradients of polarization
angle curves for the highly polarized components are concentrated near
0.0. The gradients for the low polarized components have fewer data
but are widely distributed. We therefore conclude that the highly
polarized emission tends to have a flat polarization angle curve.

\subsection{Depolarization and other properties}

There are two mechanisms for depolarization of pulsar profiles:
orthogonally polarized radiation and scattering within the
interstellar medium. Single pulse observations
\citep[e.g.][]{scr+84,scw+84} show the orthogonal modes of pulsar
emission, and highly polarized components of integrated profiles are
generally of one mode. The orthogonal modes often depolarize the
integrated profiles and lead to low polarized components, as shown for
PSRs J0814+7429 and J0922+0638 by \citet{scr+84} and
\citet{rrs+02}. PSRs J0358+5413, J1825-0935 and J2008+2513 also show
orthogonal modes and have depolarized trailing components, as listed
in Table~\ref{table:PA_gradient}.

Scattering during the propagation of pulsed emission in the
interstellar medium can also cause depolarization at the trailing
parts of profiles and result in a flat polarization angle curve
\citep{lh03}. For example, PSR J1112-6103 has a dispersion measure of
599.1 and has two highly polarized components at 3.1GHz
\citep{wj08}. But at frequencies below 1.5GHz, the effect of
scattering becomes very significant and causes depolarization in the
trailing part. The polarization angle curves are also flattened, as
indicated by the gradient values in Table~\ref{table:PA_gradient}. The
other three pulsars, PSRs J1341-6220, J1410-6132 and J1730-3350 show
similar polarization profiles due to scattering.

Millisecond pulsars exhibit highly polarized components as the normal
pulsars. PSR J0737-3039A is an orthogonal rotator and has an
inter-pulse. The leading part of the main pulse and the trailing part
of the interpulse are highly linearly polarized with a nearly constant
position angle. The gradient of polarization angle curve is near 0.0
as listed in Table~\ref{table:PA_gradient}. Orthogonal modes might
happen at the trailing part of the main pulse and the leading part of
the interpulse \citep{gkj+13}. PSR J1012+5307 is an aligned rotator
and has emission at almost all rotation phases. The trailing part of
the brightest component and all the other components are highly
linearly polarized \citep{stc99,hdv+09}. The swing of polarization
angle is nearly flat at all these phases. PSRs J1022+1001 and
J1300+1240 show similar polarization features.

\section{Theoretical explanations of highly polarized components}

It can be summarized from observations that the highly polarized
components preferably have a flat spectrum and a flat polarization
angle curve. Orthogonal modes and scattering could cause
depolarization. Millisecond pulsars exhibit similarly highly polarized
components as the normal pulsars. After we analyze literature data to
uncover these features for highly polarized components, we here carry
out numerical simulations of emission processes and propagation
effects to understand the polarization.

\subsection{A theoretical model for emission processes and propagation effects}

\begin{figure}
  \centering
  \includegraphics[angle=0, width=0.4\textwidth]{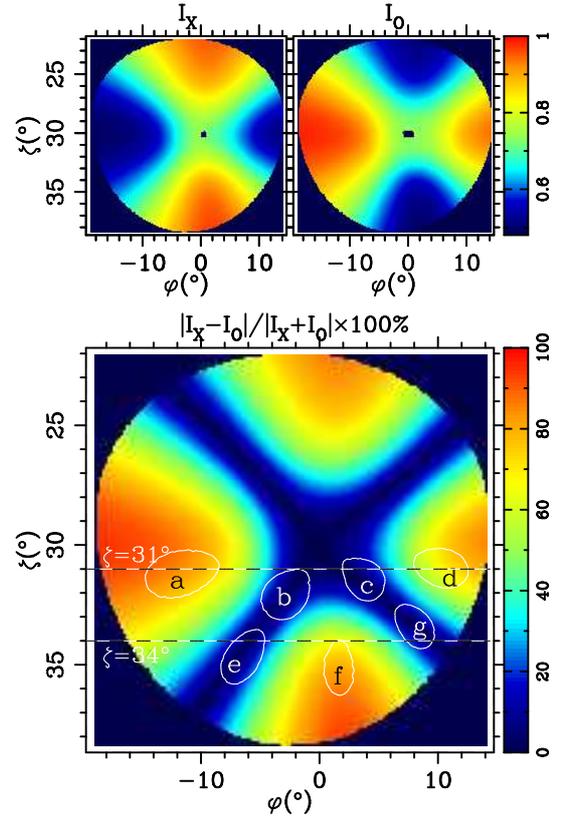}
  \caption{The distributions of wave modes and fractional linear
    polarization within simulated pulsar emission beam. The upper
    panels are plotted for the X-mode and O-mode intensities, $I_X$
    and $I_O$. The bottom panel shows the degree of linear
    polarization. Seven density patches labeled as $\it a$, $\it b$,
    $\it c$, $\it d$, $\it e$, $\it f$ and $\it g$ are shown in the
    figure and their locations are listed in
    Table~\ref{table:patch_loc}. Example sight lines at $\zeta=31^o$
    and $34^o$ from the rotation axis of a neutron star are indicated
    by the dashed lines. Other pulsar parameters used for simulations
    are the inclination angle of the magnetic axis from the rotation
    axis $\alpha=30^o$ and pulsar period $P=1s$. Relativistic
    particles are assumed to have a Lorentz factor of $\gamma=500$,
    and emit at 1.4GHz.}
  \label{fig:patch_dis}
\end{figure}

In general, pulsar magnetosphere is assumed to be an dipole,
\begin{equation}
\vecB=B_{\star}(\frac{R_{\star}}{r})^3[3\vechatr(\vechatr\cdot
\vechatm)-\vechatm],
\label{eq:staticb}
\end{equation}
here $R_{\star}$ and $B_{\star}$ represent neutron star radius and the
magnetic field on its surface, $\vechatr$ and $\vechatm$ are the unit
vectors along $\vecr$ and the magnetic dipole moment. The magnetic
axis inclines to the rotation axis by an inclination angle
$\alpha$. It rotates freely in space. Relativistic particles with a
Lorentz factor of $\gamma$ are produced by the sparking processes
above the polar cap. They stream out along the curved magnetic field
lines and co-rotate with pulsar magnetosphere. As influenced by the
perpendicular acceleration, relativistic particles will produce
curvature radiation. The radiation field $\bmath E(t)$ and its Fourier
components $\bmath E(\omega)$ can be calculated by using circular path
approximation \citep{wwh12}. Curvature radiation at a given position
of pulsar magnetosphere actually contains the contributions from all
the nearby field lines within a $1/\gamma$ cone around the tangential
direction. The polarization patterns of emission cones are further
distorted by rotation effects, as demonstrated by \citet{wwh12}.

In general, there are four wave modes (two transverse and two
longitudinal) in the plasma of pulsar magnetosphere
\citep{bp12}. Two modes are damped at large distances from the
neutron star in the magnetosphere. Only the X-mode and superluminous
O-mode, hereafter the O-mode, can escape from the magnetosphere to be
observed. Immediately after the waves are generated in the emission
region, they are coupled to the local X-mode and O-mode to propagate
outwards. Within the $1/\gamma$ emission cone, both components have
comparable intensities and propagate separately. The X-mode component
propagates in a straight line, while the O-mode component suffers
refraction \citep{ba86}. Hence, both mode components are separated
\citep{wwh14}. The detectable emission at a given position consists of
incoherent superposition of X-mode and O-mode components coming from
discrete emission regions. Both mode components experience `adiabatic
walking', wave mode coupling, and cyclotron absorption \citep{wlh10,
  bp12}.

These emission processes and propagation effects have been considered
jointly by \citet{wwh14} for four particle density models in the form
of uniformity, cone, core and patches. We demonstrated that refraction
and co-rotation significantly affect pulsar polarizations. Refraction
bends O-mode emission towards the outer part of pulsar emission beam,
and causes the separation of both modes. Co-rotation will lead to
different ratios for both modes at different parts of pulsar emission
beam. Investigations on the influences of both effects have been
extended to a wide range of frequencies, and succeeded in
demonstrating the frequency dependence of pulsar linear polarization
\citep{wwh15}.

\begin{table}
  \begin{center}
    \caption{Assumed seven density patches within a pulsar emission
      beam. Here, $\theta_i$ and $\phi_i$ represent the peak positions
      for the Gaussian density patches in the magnetic colatitude
      $\theta$ and azimuth $\phi$ directions within ranges of
      $0<\theta_i<1$ and $-180^o<\phi_i<180^o$. $\sigma_\theta$ and
      $\sigma_\phi$ represent the width of Gaussian distribution of
      the density distribution of particles. }
    \label{table:patch_loc}
    \begin{tabular}{clcrr}
      \hline
      \hline
      Index &$\theta_i$&$\phi_i(^o)$&$\sigma_{\theta}$&$\sigma_{\phi}(^o)$ \\
      \hline
      $\it a$     &  0.8   &   85   & 0.06  &  5        \\
      $\it b$     &  0.5   &   40   & 0.08  &  12       \\
      $\it c$     &  0.5   &  -55   & 0.09  &  12       \\
      $\it d$     &  0.85  &  -85   & 0.06  &  5        \\
      $\it e$     &  0.8   &   40   & 0.06  &  5        \\
      $\it f$     &  0.8   &  -10   & 0.06  &  5        \\
      $\it g$     &  0.85  &  -55   & 0.06  &  5        \\
      \hline
    \end{tabular}
  \end{center}
\end{table}

Based on our previous studies \citep{wwh12, wwh14, wwh15}, we here
simulate the curvature radiation processes and propagation effects,
but focus mainly on the distribution of highly polarized emission
regions within pulsar emission beam. Fig.~\ref{fig:patch_dis}
represents a very typical case for the distributions of wave modes and
fractional linear polarization, based on an uniform density model
demonstrated in \citet{wwh14}. It shows that the intensity
distributions for both modes are quite different. The X-mode
components, $I_X$, are stronger at the two sides of pulsar beam in the
$\zeta$ direction, as shown in the top left panel of
Fig.~\ref{fig:patch_dis}, while $I_O$ are stronger at the two sides of
pulsar beam in the $\varphi$ direction, as shown in the top right
panel of Fig.~\ref{fig:patch_dis}. Here, $\zeta$ is the sight line
angle, i.e., the angle between sight line and the rotation
axis, $\varphi$ represents the rotation phase. Depolarization is
caused by two modes. Some regions in emission beam is dominated by one
mode that can be highly polarized. The depolarization leads the
distribution of fractional linear polarization $|I_X-I_O|/|I_X+I_O|$
to be quadruple. It implies that the highly polarized emission could
be produced at four parts of pulsar emission beam, i.e., the leading
(O-mode), trailing (O-mode), top (X-mode) and bottom (X-mode) parts of
the beam.

In order to demonstrate the formation of highly polarized components,
seven density patches ($\it a$, $\it b$, $\it c$, $\it d$, $\it e$,
$\it f$ and $\it g$) are simulated as listed in
Table~\ref{table:patch_loc}. As shown in the bottom panel of
Fig.~\ref{fig:patch_dis}, patches $\it a$ and $\it d$ are dominated by
the O-mode emission, while patch $\it f$ by the X-mode. The emission
from these regions should have a large fraction of linear
polarization. However, emission from density patches $\it b$, $\it c$
and $\it e$ have both the X and O modes with comparable intensity,
hence the observed emission from these regions is depolarized.

\subsection{Polarized pulse profiles obtained by a small impact angle}

\begin{figure}
  \centering
  \includegraphics[angle=0, width=0.235\textwidth] {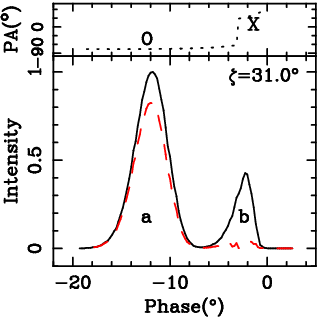}
  \includegraphics[angle=0, width=0.235\textwidth] {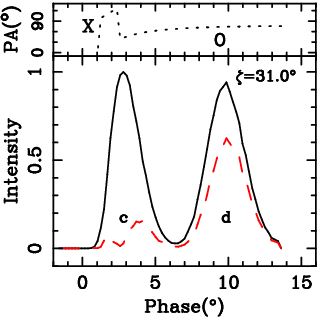}
  \caption{Pulse profiles resulting from the cut of density patches
    ($\it a$, $\it b$) and ($\it c$, $\it d$) to explain the highly
    polarized leading and trailing components, depending on the
    available density patches in the emission region. The sold lines
    represent the total intensity, the dashed and dotted lines are for
    the linear polarization and polarization angle curves. The wave
    modes are marked near the polarization angle curves.}
  \label{fig:patch_prof_s}
\end{figure}

When a sight line has a small impact angle, $\beta$, i.e.,
$\zeta-\alpha$, cutting across pulsar emission beam, it will detect
emissions from the density patches $\it a$, $\it b$, $\it c$ and $\it
d$ in Fig.~\ref{fig:patch_dis}. The resulting pulse profiles are shown
in Fig.~\ref{fig:patch_prof_s}, depending on the available density
patch combinations, for example ($\it a$, $\it b$) or ($\it c$, $\it
d$). We can conclude from simulations that:

1) The highly polarized components can be generated from the leading
(patch $\it a$) and the trailing (patch $\it d$) parts of pulsar
emission beam, both of which are dominated by the O-mode.

2) The highly polarized components have a flat polarization angle
curve, because density patches $\it a$ and $\it d$ are away from the
meridional plan of $\varphi=0^o$.

3) The low polarized components exhibit orthogonal modes, and the
emission from the X and O modes has comparable intensity. Orthogonal
mode jump happens when one mode dominates over the other, as shown by
the polarization angle curves.

In addition, the simulations predict that the highly polarized
components are more likely to be generated at the leading parts of
pulse profiles, because the highly polarized leading part of pulsar
emission beam is broader than the trailing one (see
Fig.\ref{fig:patch_dis}) due to rotation-induced asymmetry. Highly
polarized components would have a flatter spectrum than the low
polarized components, because the beam regions further away from the
magnetic axis tend to have a flat spectrum according to \citet{lm88},
while the detailed spectrum behavior is not modeled in our
simulations.

\subsection{Polarized pulse profiles obtained by a large impact angle}

\begin{figure}
  \centering
  \includegraphics[angle=0, width=0.235\textwidth] {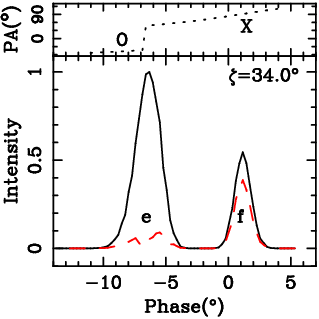}
  \includegraphics[angle=0, width=0.235\textwidth] {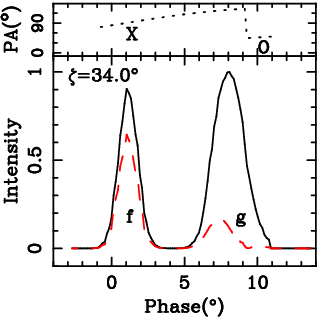}
  \includegraphics[angle=0, width=0.235\textwidth] {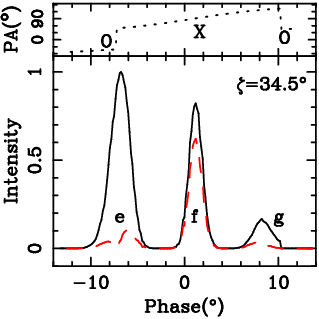}
  \caption{Same as Fig.~\ref{fig:patch_prof_s} but for density patch
    combinations of ($\it e$, $\it f$), ($\it f$, $\it g$) and ($\it
    e$, $\it f$, $\it g$).}
  \label{fig:patch_prof_l}
\end{figure}

When a sight line has a large impact angle to cut across the pulsar
emission beam, it will detect emission from density patches $\it e$,
$\it f$ and $\it g$ in Fig.~\ref{fig:patch_dis}. The resulting pulse
profiles are shown in Fig.~\ref{fig:patch_prof_l} for different patch
combinations ($\it e$, $\it f$), ($\it f$, $\it g$) or ($\it e$, $\it
f$, $\it g$) for available density distributions of particles. Highly
polarized components can appear at the leading, central or trailing
part of pulse profiles. These profiles have similar features as those
in Fig.~\ref{fig:patch_prof_s}, but differences are as following.

1) The highly polarized component from the bottom part, i.e., density
patch $\it f$, of pulsar emission beam is dominated by the X-mode,
rather than the O-mode.

2) Highly polarized component has a steeper polarization angle curve,
because the component is generated near the meridional plane, where
the polarization angle has the maximum rate of change approximating
$(d PA/d \varphi)_{\rm max}=\sin \alpha/\sin \beta$. The gradient is
inversely proportional to the impact angle $\beta$.

3) Highly polarized component may have a similar spectrum to the low
polarized components, since all components are generated at comparable
distances from the magnetic axis.

In summary, joint simulations of emission processes and propagation
effects demonstrate that highly polarized components can be produced
at the leading, central and trailing parts of pulse profiles. The
properties of emission components, polarization angle curve, mode
characteristic and spectrum, depend on pulsar geometry and the density
patches of radiating particles.

\section{Discussions and Conclusions}

In this paper, we have investigated the highly polarized components of
integrated pulse profiles observationally and theoretically. We found
from observational data that:

(i) Highly polarized components of pulsar profiles have a flatter
spectrum than the low polarized components, regardless of their
locations at the leading or trailing phase;

(ii) Highly polarized components tend to have a flat polarization
angle curve, though a small fraction of pulsars have very steep
polarization angle curves;

(iii) Highly polarized components generally have one mode, while the
low polarized components often show orthogonal modes;

(iv) Significant scattering will cause depolarization at the trailing
parts of pulse profiles and result in flat polarization angle curves;

(v) Millisecond pulsars can have highly polarized components as normal
pulsars.

We simulated emission processes and propagation effects within pulsar
magnetosphere, and found that highly polarized emission could be
produced at the leading (O-mode), trailing (O-mode), top (X-mode) and
bottom (X-mode) parts of pulsar emission beam. When a sight line cuts
across the beam with different impact angles, the detected highly
polarized components have different properties, depending on the
specific geometry and available density patches of the radiating
particles:

(i) Highly polarized component generated from the leading or trailing
part of pulsar emission beam is of the O-mode, and has a flat
polarization angle curve;

(ii) Highly polarized component generated from the top or bottom part
of pulsar emission beam is of the X-mode, and has a steep polarization
angle curve.


In the observational aspect, polarization observations at multiple
frequencies are important to reveal the frequency dependencies of
intensities and polarization degrees of the components. The
polarization observations should have higher signal to noise ratio and
time resolution. For example, PSR J1048-5832 appeared to have one
component at 1.44GHz due to limited time resolution \citep{qml+95},
but it is clearly resolved to two components by the recent
polarization observations at 1.5GHz \citep{wj08}, which show clearly
the gradient differences of polarization angle curves between the
highly polarized and low polarized components. In addition, single
pulse observations can help to identify the wave modes and
depolarization processes \citep{scr+84, scw+84}.

In the theoretical aspect, our simulations here represent the further
development of joint researches on the emission processes and
propagation effects \citep{wwh14} and focus mainly on the properties
for the highly polarized components within the wave mode separated
magnetosphere. Note, however that in our current calculations, the
magnetic field is assumed to be a rotating dipole for an empty
magnetosphere. Radiation correction is neglected, and the effect of
loaded plasma on magnetic fields is not yet incorporated.
Furthermore, the energy and density distributions of relativistic
particles are assumed to follow a simple model. Therefore, the
conclusions and predictions under these assumptions may be altered if
more complicated pulsar magnetosphere is considered.

\section*{Acknowledgements}

This work has been supported by the National Natural Science
Foundation of China (11403043, 11473034 and 11273029), and the
Strategic Priority Research Programme ``The Emergence of Cosmological
Structures'' of the Chinese Academy of Sciences (Grant
No. XDB09010200).

\bibliographystyle{mnras}
\bibliography{HiP}

%

\appendix

\section{Peak intensity ratios for highly and low polarized components}

Tables \ref{table:leading_ratio} and \ref{table:trailing_ratio} list
the peak intensity ratios, $\rm I_{HiP}/I_{LowP}$, of the highly
polarized components at the leading and trailing part of profiles with
respect to the low polarized components at a series of frequencies.

\begin{table}
\centering
\caption{The ratios for peak intensities of highly polarized leading
  components, $\rm I_{HiP}$, with respect to low polarized ones, $\rm
  I_{LowP}$, at a series of frequencies.  References are numbered in
  Table A1.}
\label{table:leading_ratio}
\tabcolsep 1.5mm
    \begin{tabular}{llrl}
      \hline
      \hline
         PSR  & Freq.(GHz) & $\rm I_{HiP}/I_{LowP}$ & References\\
      \hline
      J0358+5413  &  0.234 & $<0.05$       & 31 \\
                  &  0.325 & $<0.03$       & 77 \\
                  &  0.408 & $0.05\pm0.02$ & 16, 31 \\
                  &  0.610 & $0.17\pm0.02$ & 31 \\
                  &  0.925 & $0.36\pm0.02$ & 31 \\
                  &  1.408 & $1.00\pm0.04$ & 31 \\
                  &  1.642 & $1.33\pm0.06$ & 31 \\
                  &  1.71  & $1.77\pm0.08$ & 30 \\
                  &  1.72  & $2.13\pm0.07$ & 9, 20 \\
                  &  2.65  & $2.40\pm0.26$ & 8, 9 \\
                  &  4.85  & $9.00\pm2.26$ & 30, 34, 81 \\
                  &  8.7   & $11.00\pm5.52$& 9 \\
                  &  10.55 & $10.00\pm2.01$& 24, 30, 34 \\
                  &  32.0  & $>8.25$       & 26 \\
      \hline
      J0454+5543  &  0.234 & $0.32\pm0.05$ & 31 \\
                  &  0.408 & $0.36\pm0.02$ & 16, 31 \\
                  &  0.610 & $0.48\pm0.03$ & 31 \\
                  &  0.91  & $0.57\pm0.05$ & 31 \\
                  &  1.408 & $0.74\pm0.04$ & 30, 31 \\
                  &  1.642 & $0.70\pm0.09$ & 31 \\
                  &  1.72  & $0.81\pm0.04$ & 20 \\
                  &  4.85  & $0.83\pm0.05$ & 30, 81 \\
      \hline
      J1048-5832  &  0.692 & $0.70\pm0.06$ & 90 \\
                  &  1.369 & $1.50\pm0.09$ & 90 \\
                  &  1.5   & $1.71\pm0.06$ & 69 \\
                  &  3.0   & $3.71\pm0.27$ & 69 \\
                  &  3.1   & $4.75\pm0.91$ & 56, 90 \\
                  &  6.387 & $7.33\pm3.70$ & 90 \\
                  &  8.4   & $11.50\pm2.89$& 59, 90 \\
      \hline
      J1057-5226IP&  0.17  & $0.81\pm0.25$ & 12 \\
                  &  0.325 & $1.00\pm0.07$ & 77 \\
                  &  0.631 & $1.60\pm0.28$ & 6, 12 \\
                  &  0.64  & $1.88\pm0.40$ & 16 \\
                  &  0.692 & $1.50\pm0.27$ & 90 \\
                  &  0.95  & $2.60\pm1.95$ & 29 \\
                  &  1.369 & $2.33\pm0.21$ & 4, 72, 90 \\
                  &  1.5   & $2.15\pm0.18$ & 69 \\
                  &  3.0   & $2.78\pm0.82$ & 69, 90 \\
      \hline
      J1825-0935MP&  0.243 & $0.07\pm0.06$ & 65 \\
                  &  0.325 & $0.07\pm0.02$ & 65, 73, 77 \\
                  &  0.408 & $0.09\pm0.02$ & 31 \\
                  &  0.61  & $0.14\pm0.02$ & 31 \\
                  &  0.69  & $0.13\pm0.02$ & 63, 65 \\
                  &  0.925 & $0.17\pm0.02$ & 31 \\
                  &  0.95  & $0.19\pm0.02$ & 29 \\
                  &  1.4   & $0.20\pm0.02$ & 30, 49, 61, 65 \\
                  &  1.612 & $0.24\pm0.02$ & 7 \\
                  &  3.1   & $0.26\pm0.02$ & 61, 63, 65 \\
                  &  4.85  & $0.24\pm0.02$ & 30 \\
                  &  10.45 & $0.42\pm0.18$ & 30 \\
      \hline
      J1844+1454  &  0.243 & $<0.05$       & 65, 90 \\
                  &  0.325 & $<0.02$       & 65, 77, 90 \\
                  &  0.408 & $<0.05$       & 31   \\
                  &  0.61  & $0.25\pm0.03$ & 31   \\
                  &  0.69  & $0.33\pm0.08$ & 65, 90 \\
                  &  0.925 & $1.00\pm0.21$ & 31   \\
                  &  1.4   & $1.15\pm0.09$ & 17, 31, 38 \\
                  &        &               & 54, 65, 74 \\
                  &  1.642 & $1.18\pm0.23$ & 31   \\
                  &  3.1   & $2.00\pm0.67$ & 65, 90 \\
      \hline
    \end{tabular}
\parbox{85mm}{Note: The highly polarized components of PSR J0358+5413
  and J1844+1454 are confused with the corresponding low polarized
  components at frequencies smaller than 0.408 and 0.61GHz.}
\end{table}

\begin{table}
\centering
\caption{Same as Table~\ref{table:leading_ratio} but for highly
  polarized trailing components. }
\label{table:trailing_ratio}
\tabcolsep 1.5mm
    \begin{tabular}{llrl}
      \hline
      \hline
         PSR  & Freq.(GHz) &$\rm I_{HiP}/I_{LowP}$& References\\
      \hline
      J0601-0527  &  0.243 & $0.34\pm0.14$ & 90    \\
                  &  0.325 & $0.42\pm0.05$ & 90    \\
                  &  0.408 & $0.45\pm0.08$ & 31    \\
                  &  0.61  & $0.50\pm0.03$ & 31    \\
                  &  0.692 & $0.66\pm0.09$ & 90    \\
                  &  0.925 & $0.63\pm0.19$ & 31    \\
                  &  1.408 & $0.95\pm0.07$ & 31, 54, 90  \\
                  &  1.5   & $1.02\pm0.05$ & 69    \\
                  &  1.642 & $1.05\pm0.19$ & 31    \\
                  &  1.72  & $1.30\pm0.15$ & 20    \\
                  &  3.068 & $1.79\pm0.26$ & 90    \\
                  &  4.85  & $2.12\pm0.52$ & 34    \\
                  &  6.2   & $2.86\pm1.41$ & 90    \\
      \hline
      J0922+0638  &  0.243 & $0.77\pm0.05$ & 65, 90 \\
                  &  0.322 & $1.0\pm0.07$  & 62, 65, 90   \\
                  &  0.43  & $1.1\pm0.05$  & 19    \\
                  &  0.690 & $1.59\pm0.13$ & 65, 90\\
      \hline
      J1539-5626  &  0.692 & $0.45\pm0.37$ & 90    \\
                  &  1.377 & $0.60\pm0.08$ & 90    \\
                  &  1.5   & $0.58\pm0.03$ & 69    \\
                  &  3.1   & $0.78\pm0.06$ & 56, 69, 90 \\
                  &  6.2   & $1.77\pm0.65$ & 90    \\
                  &  8.356 & $1.95\pm0.58$ & 59, 90\\
     \hline
      J2225+6535  &  0.15  & $0.07\pm0.06$ & 88    \\
                  &  0.234 & $0.20\pm0.10$ & 31    \\
                  &  0.325 & $0.28\pm0.10$ & 77    \\
                  &  0.408 & $0.37\pm0.14$ & 16, 31\\
                  &  0.61  & $0.57\pm0.05$ & 31    \\
                  &  0.925 & $1.11\pm0.17$ & 31    \\
                  &  1.408 & $1.33\pm0.07$ & 31    \\
                  &  1.5   & $1.39\pm0.07$ & 86    \\
                  &  1.642 & $1.43\pm0.19$ & 31    \\
                  &  1.7   & $1.38\pm0.21$ & 9     \\
      \hline
    \end{tabular}
\parbox{85mm}{ Note: The highly polarized components of PSR J0922+0638
  are confused with the low polarized components at frequencies larger
  than 0.69GHz.}
\end{table}

\label{lastpage}

\end{document}